\documentclass[aps,prb,preprint,groupedaddress]{revtex4-1} 
\usepackage{graphicx}
\usepackage{dcolumn} 
\usepackage{bm} 
\usepackage{color}
\usepackage{amsmath}
\usepackage{tikz}
\usetikzlibrary{arrows,shapes}
\begin{document}

\title{Excitonic properties of F-centers in $\alpha$-alumina from First Principles Calculation}

\author{Tathagata Biswas} 
\affiliation{Center for Condensed Matter Theory, Department of Physics, Indian Institute of Science, Bangalore, 560012}  
\author{Manish Jain}
\email{mjain@iisc.ac.in} 
\affiliation{Center for Condensed Matter Theory, Department of Physics, Indian Institute of Science, Bangalore, 560012}

\date{\today}

\begin{abstract} 
We use state-of-the art GW-BSE formalism to study electronic structure and
optical properties of oxygen vacancies in $\alpha$-alumina.  Many body
perturbation theory within GW approximation in recent years have been used
extensively to study excited state properties of a wide range of systems.
Moreover, solving Bethe-Salpeter equation (BSE) enable us to capture excitonic
effects in a material. We compute the charge transition levels (CTLs) for oxygen
vacancies using DFT+GW formalism. We propose an alternative approach to
calculate these CTLs, which provides a more efficient way to perform
electrostatic correction required because of finite supercell sizes and
periodic boundary condition used in first principles calculations. We find that
oxygen vacancy in this material has deep donor levels, (+2/+1) at 2.5 eV and a
(+1/0) level at 3.8 eV above the VBM. We also study F-center absorption and
emission processes using constrained--DFT and BSE. Our calculated absorption
and emission energies are in excellent agreement with experimental results.  
\end{abstract}

\keywords{Density functional theory, Many body perturbation theory, Defects} 

\maketitle

\section{Introduction}

Aluminium oxide with corundum structure ($\alpha$-alumina) is perhaps one the
most used structural ceramic in the world. The application of this material
ranges from high-temperature structural ceramics, abrasive, dielectric
insulators, catalyst to optical devices
\cite{pearson1977alumina,auerkari1996mechanical,dragic2012sapphire,zahler2016mobile}.
$\alpha$-alumina has been used not only in traditional fields such as cutting
tools in industries, substrates for the growth of thin metal, semiconductor,
and insulator films but also in exciting new applications such as strong
durable optical fibres or scratch resistant screens on mobile electronics
devices \cite{dragic2012sapphire,zahler2016mobile}. This is because of it's
unusual combination of mechanical, chemical and electronic properties
\cite{auerkari1996mechanical}.  However, all these properties can alter
dramatically as we consider point defects or impurities in an otherwise perfect
crystal \cite{heuer1980plastic,cannon1980plastic}. For instance anionic vacancy
in a crystal which is filled by one or more electrons (F-center or color
center) has been found to absorb light in the visible spectrum making a wide
band gap transparent material, colored \cite{evans1991optical}. As a result,
not only aluminium oxide as a material but also defects in aluminium oxide have
been focus of extensive theoretical and experimental study for past few decades
\cite{evans1991optical,matsunaga2003first,french1990electronic,french1994interband,marinopoulos2011local,choi2013native}.
Oxygen related defects especially oxygen vacancies, are known to be a common
defect in most oxides \cite{campbell2005oxygen}.  However, the abundance of
experimental data available in this field has not been complemented with
theoretical studies  to provide meaningful insight into these defects. 

Thermodynamic CTLs provide useful information from an electronic or
optoelectronic application point of view. CTL not only describes whether the
defect state is going to act as a donor or acceptor but also reports if it is a
shallow one ($\sim$ $k_BT$ from bands edges) or lies deep inside the band gap
\cite{freysoldt2014first}. Shallow levels being close to VBM or CBM can produce
electrons in conduction band or holes in valence band though thermal excitation
alone and therefore results in controlled $n$-type or $p$-type conductivity.
However, sometimes shallow levels are not technologically desirable. If
unintentional dopants (impurities) introduces a shallow level, it can results
in a reduction of doping efficiency. A deep level can also affect device
performance and may reduce device lifetimes of electronic or optoelectronic
devices. As it can provide uncontrolled radiative or nonradiative recombination
channels.  It should be noted that in recent years, deep levels have been used
constructively as well. For example, one can use them to pin the Fermi level in
an energy region far from the band edges
\cite{robertson2007fermi,chadi1997fermi}. Moreover, deep levels have also been
used as single spin centers for quantum computing in systems like
nitrogen-vacancy (NV) center in diamond \cite{weber2010quantum}.

\section{Computational Methods}

We perform all the DFT calculations using generalized gradient approximation 
(GGA) \cite{perdew1996generalized} and norm-conserving pseudopotentials
\cite{troullier1991efficient} as implemented in the QUANTUM ESPRESSO package
\cite{giannozzi2009quantum}. The wavefunctions are expanded in plane waves
with energy upto 75 Ry. The calculations for perfect crystal have been done using
a unit cell containing 30 atoms, with a 4$\times$4$\times$2 k-point sampling of
the Brillouin zone. For the calculations with oxygen vacancies in
various charge states, we use  2$\times$2$\times$1 and 3$\times$3$\times$1
supercells, containing 120 and 270 atoms respectively, to exclude any short
range defect-defect interactions. The Brillouin zone for 2$\times$2$\times$1 and
3$\times$3$\times$1 supercells, was sampled using 2$\times$2$\times$2 and
$\Gamma$-point respectively. We simulate different charge states ($q = 0, 1,
2$) of the oxygen vacancy. 

Quasiparticle and optical properties are calculated within the GW-BSE formalism
as implemented in the BerkeleyGW package \cite{deslippe2012berkeleygw}. To
compute the quasiparticle energies (E$^{QP}$) we solve the Dyson equation
\cite{hedin1970effects}, where the self energy operator ($\Sigma(E)$) has been
calculated within G$_0$W$_0$ approximation. The dielectric matrix ($\epsilon$)
is calculated within the random phase approximation (RPA) and expressed in a
plane-wave basis with plane wave energies upto 25 Ry. The matrix is calculated
at zero frequency and extended to finite frequencies using generalized plasmon
pole model proposed by Hybertsen and Louie \cite{hybertsen1986electron}. In
case of 30 atom unit cell, we include 1000 bands while performing sum over
unoccupied states involved in the dielectric matrix and self-energy
calculations. For 2$\times$2$\times$1 and 3$\times$3$\times$1 supercells, we
increase the number of bands to 4000 and 9000 respectively. To study
electron-hole interaction and excitonic effects we solve the Bethe-Salpeter
equation \cite{rohlfing2000electron}. Electron-hole interaction kernel for BSE
calculation is computed with 44 valence and 11 conduction bands. The
calculated kernel is extrapolated from a 4$\times$4$\times$2 to a finer
10$\times$10$\times$5 k-point sampling of the Brillouin zone. These parameters
are sufficient to obtain a converged optical spectrum for energies upto $\sim$
5 eV from absorption edge.  

The formation energy of a point defect, $\mathrm{E^f_q({\bf R})[\epsilon_F]}$,
in charge state $q$, with all the atoms at coordinates $\left\{{\bf R}\right\}$
and the chemical potential of the electron (Fermi level)
$\mathrm{\epsilon_{F}}$, can be defined as \cite{freysoldt2014first},

\begin{equation}
\label{eqn1} 
\mathrm{E^f_q({\bf{R}})[\epsilon_{F}]=E^{def}_q({\bf{R}})-[E^{perf}+ \sum_i n_i \mu_i]+[\epsilon_F+E_v] q }
\end{equation}

where $\mathrm{E^{def}_q({\bf{R}})}$ is the total energy of the defect
supercell in charge state q with all the atoms at coordinates $\left\{{\bf
R}\right\}$ and $\mathrm{E^{perf}}$ is the total energy of the perfect
supercell (without any defects). $\mathrm{n_{i}}$ refers to the number of atoms
removed ($\mathrm{n_{i}}<0$) from or added to ($\mathrm{n_{i}}>0$) the perfect
supercell to make the defect supercell. The removed/added atoms are exchanged
from a bath with chemical potential $\mathrm{\mu_{i}}$. It should be noted that
we have defined $\mathrm{\epsilon_{F}}$ with respect to valence band maxima
($\mathrm{E_v}$). 

\begin{figure}[h!]
\setlength{\unitlength}{0.1\textwidth}
\begin{picture}(4,4)
\put(-2,0.2){\includegraphics[height=5cm] {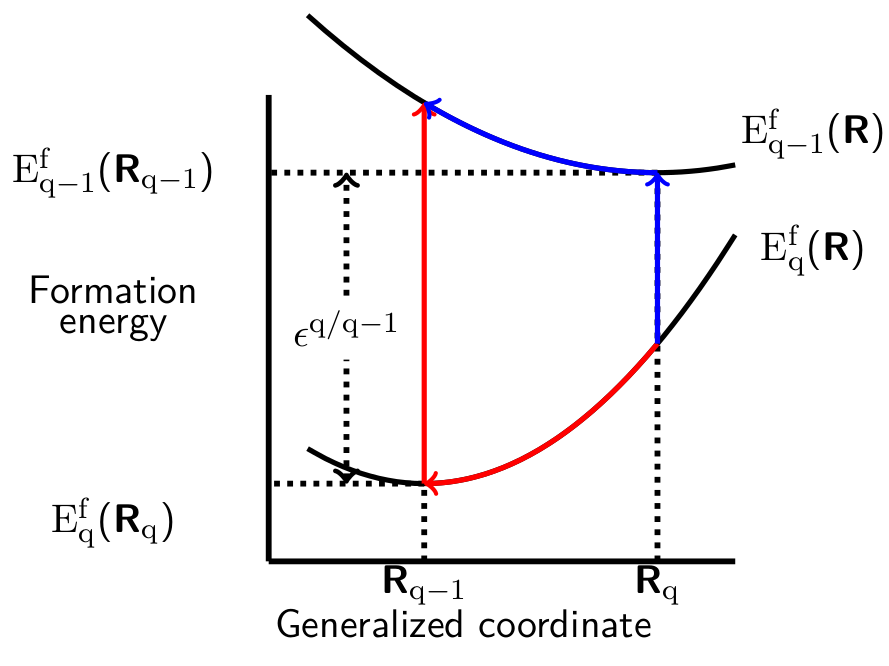}}
\put(2.5,0.2){\includegraphics[height=5cm] {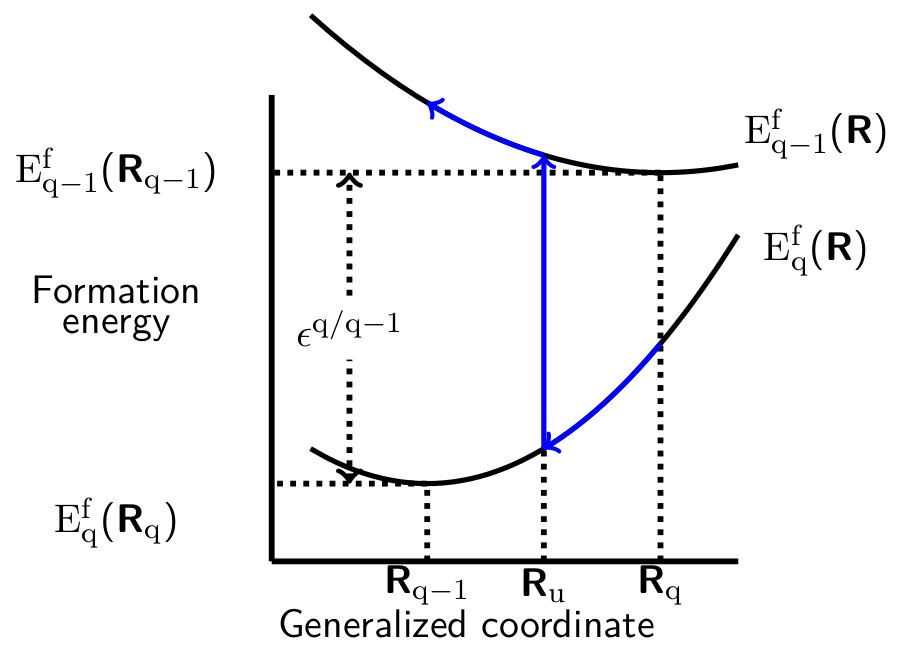}}
\put(0,0){(a)}
\put(4.5,0){(b)}
\end{picture}
\caption{Paths that can take the system being at equilibrium with defect at
charge state $q$ to one where it is at charge state $q-1$. Paths that is
equivalent to using (a) Eq.~\ref{eqn3} (red) Eq.~\ref{eqn4} (blue) and (b)
Eq.~\ref{eqn5} have been shown.}
\label{fig1}
\end{figure}

Thermodynamic charge transition level (CTL) $\mathrm{\epsilon^{q/q-1}}$, is
defined as the value of the electron chemical potential
($\mathrm{\epsilon_{F}}$) at which the charge state of the defect changes from
$q$ to $q-1$. It can be written in terms of formation energies as
\cite{freysoldt2014first},

\begin{equation}
\label{eqn2}
\mathrm{\epsilon^{q/q-1}=E^f_{q-1}({\bf{R}}_{q-1})[\epsilon_{F}=0]-E^f_{q}({\bf{R}}_{q})[\epsilon_{F}=0]}
\end{equation}

where $\mathrm{\left\{{\bf R}_q\right\}}$ and $\mathrm{\left\{{\bf
R}_{q-1}\right\}}$ denote the equilibrium structures with defect in charge
states $q$ and $q-1$ respectively.

Within standard DFT, one can obtain the formation energies in equilibrium
configurations of respective charge states and thus calculate
$\mathrm{\epsilon^{q/q-1}}$. However, as we have mentioned earlier that DFT is
a theory for ground state properties. Calculating CTL involves addition or
removal of one or multiple electrons, therefore can not be accomplished with
DFT formalism alone. To overcome this problem we have used a more sophisticated
DFT+GW formalism \cite{jain2011quasiparticle,rinke2009defect,PhysRevLett.97.226401}. 
In this method we write the CTL as,

\begin{equation}
\label{eqn3}
\begin{split}
\mathrm{\epsilon^{q/q-1}} & = \mathrm{[E^f_{q-1}({\bf{R}}_{q-1})-E^f_{q-1}({\bf{R}}_{q})]
                              +[E^f_{q-1}({\bf{R}}_{q})-E^f_{q}({\bf{R}}_{q})]} \\  
                              & \equiv \mathrm{E_{q-1}^{relax}[{\bf R}_q]+E^{QP}(\bf{R}_q)}
\end{split}
\end{equation}

The first term ($\mathrm{E_{q-1}^{relax}[{\bf R}_q]}$) on the right hand side
of Eq.~\ref{eqn3} is a relaxation energy. For a system containing a defect in
charge state $q-1$, $\mathrm{E_{q-1}^{relax}[{\bf R}_q]}$ is the total energy
difference between a structure with all the atoms at coordinates
$\mathrm{\left\{{\bf R}_q\right\}}$ and the equilibrium structure (in this case
$\mathrm{\left\{{\bf R}_{q-1}\right\}}$). The relaxation energy can be computed
accurately within DFT. The second term on the right hand side
$\mathrm{E^{QP}({\bf R}_q)}$ of Eqn.~\ref{eqn3}, involves a change in the
electron number as the defect charge changes from $q$ to $q-1$. This is the
electron affinity (EA) of the system containing a defect in charge state $q$
with all the atoms at $\mathrm{\left\{{\bf R}_q\right\}}$. This quasiparticle
energy can be computed accurately using the GW formalism. Alternatively, we can
write, 

\begin{equation}
\begin{split}
\label{eqn4}
\mathrm{\epsilon^{q/q-1}} & = \mathrm{[E^f_{q-1}({\bf{R}}_{q-1})-E^f_{q}({\bf{R}}_{q-1})]
                              +[E^f_{q}({\bf{R}}_{q-1})-E^f_{q}({\bf{R}}_{q})]} \\
                              & \equiv \mathrm{E^{QP}({\bf R}_{q-1}) - E_{q}^{relax}[{\bf R}_{q-1}]} 
\end{split}
\end{equation}

where $\mathrm{E_{q}^{relax}[{\bf R}_{q-1}]}$ is defined in the same way as
$\mathrm{E_{q-1}^{relax}[{\bf R}_q]}$ in Eq.~\ref{eqn3} and
$\mathrm{E^{QP}({\bf R}_{q-1})}$ is the ionization potential (IP) of the system
containing a defect in charge state $q-1$ with all the atoms at
$\mathrm{\left\{{\bf R}_{q-1}\right\}}$. In Fig~\ref{fig1}(a) we show two
paths that can take the system being at equilibrium with defect at charge state
$q$ to one where it is at equilibrium with charge state $q-1$. Using
Eq.~\ref{eqn3} or Eq.~\ref{eqn4} to compute $\mathrm{\epsilon^{q/q-1}}$ is
equivalent to taking the red or the blue path shown in Fig.~\ref{fig1}(a).
Following either of the two paths one should get the same Theromodynamic CTLs.
This DFT+GW formalism should eliminate the errors in calculating CTLs within 
standard DFT.   

First principles calculations on defects often use periodic boundary
condition (PBC) with a finite supercell size. While the supercell can be 
made to be large enough to eliminate any short range interaction between a
defect and it's periodic image, the long range interactions, such as Coulomb
interaction between charged defects, can not be eliminated entirely. To tackle
this issue, we break down the defect-defect interaction within PBC into a
short-range part such as quantum-mechanical (overlap of the wave functions),
and a long range part such as elastic or electrostatic interactions.  In
practice, we work with a supercell size that is large enough to eliminate any
short-range defect-defect interactions. For the long-range electrostatic
and elastic interactions we use a correction/extrapolation scheme. As elastic
interactions decay faster (1/L$^3$) than electrostatic interactions (1/L),
typically, one worries more about the later. In this study, we use FNV
electrostatic correction scheme \cite{freysoldt2009fully} as implemented in
CoFFEE package \cite{naik2018coffee}.

While calculating relaxation energies using DFT formalism one may be tempted to
assume that the electrostatic correction involved is zero as the total energy
corrections for two formation energies with defect in the same charge state
cancel each other. This is not entirely true, as the lattice screening in case
of the two structures are different. The lattice relaxation in the two
structures are as if they are screening different defect charges (as they
corresponds to $\mathrm{\left\{{\bf R}_q\right\}}$ and $\mathrm{\left\{{\bf
R}_{q-1}\right\}}$).  Moreover the choice of dielectric constant, while
performing eigenvalue correction to compute quasiparticle excitation energies
for charged defects is ambiguous if one uses Eq.~\ref{eqn3} or Eq.~\ref{eqn4}.
This is because an unknown fraction of lattice dielectric constant is included
owing to small supercell sizes. Using either the electronic dielectric constant
($\epsilon_{elec}$) or the full dielectric constant ($\epsilon_{total}$) is not
expected to provide accurate electrostatic correction. We propose an
alternative path to calculate CTLs which does not suffer from this issue. The
CTL can be calculated using the path shown in Fig.~\ref{fig1}(b). The CTL using
this path can be written as:

\begin{equation}
\label{eqn5}
\begin{split}
\mathrm{\epsilon^{q/q-1}}  & = \mathrm {[E^f_{q-1}({\bf{R}}_{q-1})-E^f_{q-1}({\bf{R}}_{u})]} 
                            + \mathrm{ {[E^f_{q-1}({\bf{R}}_{u})-E^f_{q}({\bf{R}}_{u})]} + { [E^f_{q}({\bf{R}}_{u})-E^f_{q}({\bf{R}}_{q})]}}\\ 
                           & = \mathrm{E^{relax}_{q-1}[{\bf R}_u] + {E_u^{QP}} - {E^{relax}_q}[{\bf R}_u] } 
\end{split}
\end{equation}

where $\mathrm{\left\{{\bf R}_u\right\}}$  denotes the unrelaxed structure,
which can be obtained by removing an atom from perfect supercell without
changing positions of any other atoms. We compute $\mathrm{E^{relax}_{q}[{\bf
R}_u]}$ ($\mathrm{E^{relax}_{q-1}[{\bf R}_u]}$) using DFT. These relaxation
energies correspond to the relaxation of defect supercell with the charge
state $q$ or $q-1$ from $\mathrm{\left\{{\bf R}_u\right\}}$  to corresponding
equilibrium structures. Quasiparticle excitation energy in Eq.~\ref{eqn5}
($\mathrm{E_u^{QP}}$) can be computed as either IP of the unrelaxed system
with defect charge state $q-1$ (P1) or EA of the same structure with defect charge
state $q$ (P2). The choice of path shown in Fig.~\ref{fig1}(b) is motivated by the
fact that the dielectric constant required for electrostatic correction of the 
$\mathrm{E_u^{QP}}$ term is well defined ($\epsilon_{elec}$).

To relax the structure for the triplet excited state of F-center
we use constrained-DFT calculation. We perform a spin polarized DFT calculation
and fix the occupations in such a way that one electron from defect state gets
promoted to the conduction band with it's spin flipped. Therefore the system  
remains neutral and mimics a triplet excited state. We show that using this
method we can calculate relaxation energies of triplet excited state, which
combined with BSE calculation provide very good agreement with experimental 
photoluminesence emission results. We find that the geometries obtained within 
such a relaxation is very close to the F$^{+}$ gerometry.

\section{Results and Discussion}

$\alpha$-alumina has a hexagonal unit cell (space group R$\bar{3}$C) containing
30 atoms or 6 formula units. It can also be constructed with a smaller
rombohedral unit cell containing 10 atoms (2 formula units). We have used
hexagonal unit cell of $\alpha$-alumina in all our calculations.

\begin{table*}[h!]
\begin{tabular}{ |m{2.5cm} | m{2.5cm}| m{2.5cm}| m{1.5cm}| m{1.5cm}| m{1.5cm}| }
\hline
 parameters &  this work (LDA) & this work (GGA) & Ref. \cite{marinopoulos2011local} (LDA)& Ref.\cite{matsunaga2003first} (GGA) & Expt. \cite{lee1985structural}  \\
\hline
\hline
  a in \AA & 4.68 & 4.81 & 4.69 &4.82 & 4.76 \\
\hline
  c in \AA & 12.45 & 13.13 & 12.79 & 13.16 & 12.99 \\
\hline
 $\mathrm {E_g}$ in eV & 6.75 & 5.82 & 6.72 & 5.82 & 9.4 \\
\hline
\end{tabular}
\label{tab1}
\caption{\label{tab1} Comparison between lattice parameters and band gap values from our calculations with literature}
\end{table*}

\begin{figure}[h!]
\setlength{\unitlength}{0.1\textwidth}
\begin{picture}(5,5)
\put(0,0){\includegraphics[height=6cm] {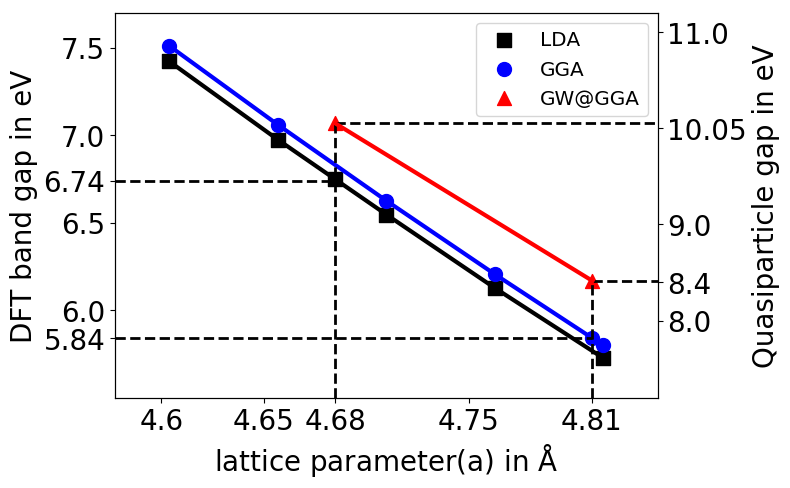}}
\end{picture}
\caption{Variation of band gap with lattice parameter a calculated using both
GGA and LDA, the equilibrium lattice parameter in both case and the
corresponding band gap has been marked.  In these calculations the c/a ratio
was fixed to the equilibrium value.}
\label{fig2}
\end{figure}

The lattice parameters of $\alpha$-alumina that we found in our calculation are
in very good agreement with previous DFT calculations
\cite{matsunaga2003first,marinopoulos2011local} and experimental values
\cite{lee1985structural} as shown in Table.~\ref{tab1}.  Our calculations show
that the band gap of this material has a strong lattice parameter dependence.
We study the band gap of $\alpha$-alumina as we change the in-plane lattice
parameter (a) by keeping the c/a ratio fixed using both LDA (PZ) and GGA (PBE)
and show the results in Fig.~\ref{fig2}.  The band gap of this material
increases almost linearly as we decrease the in-plane lattice parameter (a). Due to
overbinding effect of LDA, the band gap of the equlibrium structure within LDA
is overestimated comapared to that found within GGA (using it's equlibrium
structure). As we can see in Fig.~\ref{fig2} the difference between GGA and
LDA band gap value at there respective equilibrium lattice parameter is
$\sim{1}$eV. This difference is almost entirely because of the difference in
the lattice parameters in these two calculations.  These results explain the
large variation in the values of band gap reported in literature
\cite{matsunaga2003first,marinopoulos2011local,french1990electronic,french1994interband}.
Fig.~\ref{fig2} also shows our calculation of the GW quasiparticle gap starting
from a DFT calculation with GGA. The GW gap also varies significantly as a
function of in-plane lattice parameter inheriting the dependence from the DFT
starting point.

\begin{table}[h!]
\begin{tabular}{ | c| c| c| c| c|}
\hline
Results (in eV) & this work  & Reference \cite{marinopoulos2011local} & Expt. \cite{french1994interband} at 0K  & Expt. \cite{french1994interband} at 300K \\
\hline
\hline
 $\mathrm {E_g}$ & 9.1 & 9.36 & 9.57 & 9.31 \\
\hline
 $\mathrm {E_{EBE}}$ & 0.37 & 0.4 & 0.13 & 0.15 \\
\hline
\end{tabular}
\caption{ \label{tab2} Quasiparticle band gap and exciton binding energy
obtained from our calculation compared with values reported in literature} 
\end{table}

Table.~\ref{tab2} shows the quasiparticle band gap ($\mathrm {E_{g}}$) and
exciton binding energy ($\mathrm {E_{EBE}}$) obtained from our calculation and
compares them with values reported in the literature. Our calculated quasiparticle
band gap is in reasonable agreement with previous GW calculations done on this
material \cite{marinopoulos2011local}. The difference between the two calculations
can be attributed to the difference in the lattice parameters used. Previous study 
\cite{marinopoulos2011local} used lattice parameter values obtained from DFT
calculation using LDA, which is expected to result in a larger DFT as well as
quasiparticle gap compared to one obtained using experimental lattice
parameters as we have shown earlier (Fig.~\ref{fig2}).  However, the exciton
binding energy reported in that \cite{marinopoulos2011local} study is very
close to the one we obtain with using experimental lattice parameters. This
suggests a weaker lattice parameter dependence of $\mathrm {E_{EBE}}$ compared
to $\mathrm {E_{g}}$. We also find that GW quasiparticle gap using experimental
lattice constants (9.1 eV) is in close agreement with the experimental gap at
room temperature (9.31 eV) \cite{french1994interband}. This is because we have
used we have used the experimental lattice parameters obtained at room
temperature in our studies \cite{lee1985structural}.

\begin{figure}[h!]
\hspace{0cm}
\setlength{\unitlength}{0.1\textwidth}
    \begin{tikzpicture}[scale=0.68]

      \tikzset{ >=stealth',pil/.style={->,thick,shorten <=2pt,shorten >=2pt}}
      \draw[thick] (0,0) rectangle (5.0,6.2);

      \draw[thick] (1.0,0) edge[pil] (1.0,1.74);
      \draw[thick] (0.5,1.74) edge (1.5,1.74);
      \draw (0.8,1.74) circle (0.15) [thick];
      \draw (1.2,1.74)  circle (0.15) [thick];
      \node[fill=white] at (1.0,0.8) {1.74};
      \node at (1.0,-0.4) {V$^{+2}_{\textrm{O}}$};

      \draw[thick] (2.5,0) edge[pil] (2.5,1.5);
      \draw[thick] (2.2,1.5) edge (2.8,1.5);
      \filldraw (2.5,1.5) circle (0.15) [thick];
      \node[fill=white] at (2.4,0.8) {2.40};

      \draw[thick] (3,0,0) edge[pil] (3,3.01);
      \draw[thick] (2.2,3.01) edge (2.8,3.01);
      \node [fill=white] at (3,2.1) {3.01};
      \draw (2.5,3.01) circle (0.15) [thick];
      \node at (2.5,-0.4) {V$^{+1}_{\textrm{O}}$};

      \draw[thick] (4.0,0) edge[pil] (4.0,2.29);
      \draw[thick] (3.5,2.29) edge (4.5,2.29);
      \filldraw (3.8,2.29) circle (0.15) [thick];
      \filldraw (4.2,2.29)  circle (0.15) [thick];
      \node[fill=white] at (4.0,1.5) {2.29};
      \node at (4.0,-0.4) {V$^{0}_{\textrm{O}}$};

      \foreach \y in {0,1,2,3,4,5,6} {
        \node at (-0.3,\y + 0.1) {\y};
        \draw[thick] (0.0,\y) edge (0.2,\y);
      }
      \node [rotate=90] at (-0.75,2) {Energy (eV)};
      \node [font=\Large] at (2.5,-1.5) {DFT};

    \end{tikzpicture}
    \begin{tikzpicture}[scale=0.68]
      \tikzset{ >=stealth',pil/.style={->,thick,shorten <=2pt,shorten >=2pt}}
      \draw[thick] (0,0) rectangle (5.0,9.1);

      \draw[thick] (1.0,0) edge[pil] (1.0,4.35);
      \draw[thick] (0.5,4.35) edge (1.5,4.35);
      \draw (0.8,4.35) circle (0.15) [thick];
      \draw (1.2,4.35)  circle (0.15) [thick];
      \node[fill=white] at (1.0,3) {4.35};
      \node at (1.0,-0.4) {V$^{+2}_{\textrm{O}}$};

      \draw[thick] (2.5,0) edge[pil] (2.5,1.85);
      \draw[thick] (2.2,1.85) edge (2.8,1.85);
      \filldraw (2.5,1.85) circle (0.15) [thick];
      \node[fill=white] at (2.4,1.) {1.85};
      \draw[thick] (3,0,0) edge[pil] (3,5.15);
      \draw[thick] (2.2,5.15) edge (2.8,5.15);
      \draw (2.5,5.15) circle (0.15) [thick];
      \node[fill=white] at (3,3.5) {5.15};
      \node at (2.5,-0.4) {V$^{+1}_{\textrm{O}}$};

      \draw[thick] (4.0,0) edge[pil] (4.0,2.55);
      \draw[thick] (3.5,2.55) edge (4.5,2.55);
      \filldraw (3.8,2.55) circle (0.15) [thick];
      \filldraw (4.2,2.55)  circle (0.15) [thick];
      \node[fill=white] at (4.0,1.6) {2.55};
      \node at (4.0,-0.4) {V$^{0}_{\textrm{O}}$};

      \foreach \y in {0,1,2,3,4,5,6,7,8,9} {
        \node at (-0.3,\y + 0.1) {\y};
        \draw[thick] (0.0,\y) edge (0.2,\y);
      }
      \node [rotate=90] at (-0.7,3) {Energy (eV)};
      \node [font=\Large] at (2.5,-1.5) {GW};
    \end{tikzpicture}

\caption{Defect level positions and there occupations for different charge states
from DFT as well as GW calculation}
\label{fig3}
\end{figure}
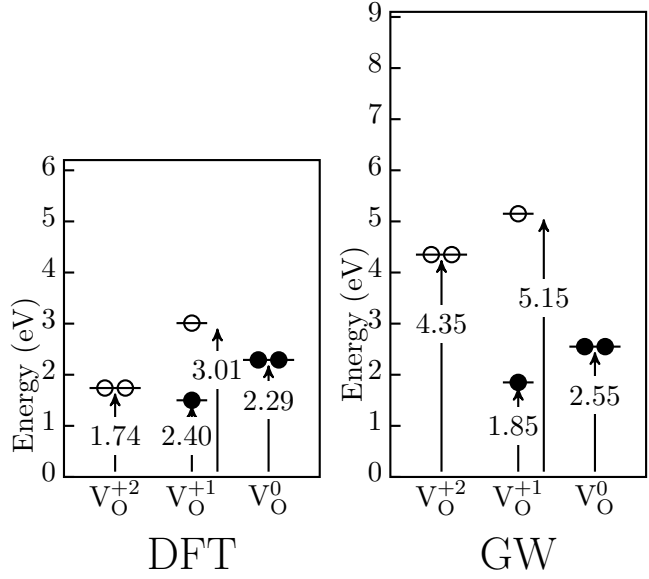

Fig.~\ref{fig3} shows the defect levels inside the band gap originated due to
an oxygen vacancy, calculated from DFT as well GW calculations performed on a
3$\times$3$\times$1 supercell.  We show the position as well occupation of
these levels for three possible charge states, neutral (V$^{0}_{\textrm{O}}$)
+1 (V$^{+1}_{\textrm{O}}$), +2 (V$^{+2}_{\textrm{O}}$).  For
V$^{0}_{\textrm{O}}$ we find one doubly occupied spin degenerate level inside
the gap.  As we remove one electron from the neutral defect the system becomes spin
polarized with only one of the spin-split defect level occupied. Removing one more
electron from the defect results in an unoccupied spin degenerate level inside
the gap. The defect levels obtained from DFT calculations do not move rigidly as we
include quasiparticle corrections. The occupied defect levels move higher in
energy by only $\sim$ 0.3 eV, whereas the unoccupied levels move by 2.1 eV 
and 2.6 eV for V$^{+1}_{\textrm{O}}$ and V$^{+2}_{\textrm{O}}$ respectively.   
In the +2 and +1 charge state the four Al atoms next to oxygen vacancy on an
average relax outward by $\sim$ 0.23 ${\mathrm \AA}$ and $\sim$ 0.1 ${\mathrm
\AA}$ respectively, whereas in the neutral charge state, the Al atoms relax
inward by $\sim$ 0.05 ${\mathrm \AA}$.

\begin{table}[h!]
\begin{center}
\begin{tabular}{ | c| c| c| c| c| c| c| c|}
\hline
CT levels & P1$^{we}$ & P2$^{we}$ & $\Delta^{we}$& P1$^{ec}$ & P2$^{ec}$ & $\Delta^{ec}$ & Mean$^{ec}$ \\
\hline
\hline
 $\mathrm{\epsilon^{+1/0}_{120}}$  & 4.87 & 3.58 & 1.3  & 3.66 & 3.58 & 0.08 &  3.62\\
\hline
 $\mathrm{\epsilon^{+1/0}_{270}}$  & 4.86 & 3.77 & 1.09 & 3.96 & 3.77 & 0.19 &  3.86\\
\hline
 $\mathrm{\epsilon^{+2/+1}_{120}}$ & 5.14 & 3.72 & 1.42 & 2.79 & 2.51 & 0.28 &  2.65\\
\hline
 $\mathrm{\epsilon^{+2/+1}_{270}}$ & 5.04 & 3.84 & 1.20 & 3.23 & 2.94 & 0.29 &  3.08\\
\hline
\end{tabular}
\caption{ \label{tab3} Charge transition levels calculated with two different cell
sizes using Eq.~\ref{eqn3} (P1) and Eq.~\ref{eqn4} (P2). We report the values obtained
with (ec) or without electrostatic corrections (we) and the differences between them ($\Delta$).}
\end{center}
\end{table}

In Table.~\ref{tab2} we show thermodynamic charge transition levels calculated
using Eq.~\ref{eqn3} (P1) and Eq.~\ref{eqn4} (P2) with two different cell
sizes, 2$\times$2$\times$1 and 3$\times$3$\times$1. We report the values
obtained with (ec) as well as without electrostatic corrections (we) to show
the importance of electrostatic correction while calculating CTL. We have used
$\epsilon=\epsilon_{elec}=3.1$ while performing the electrostatic corrections,
which we have taken from the literature \cite{kumagai2014electrostatics}. One
can see from Table.~\ref{tab2} the CTL calculated without electrostatic
correction following P1 and P2 are different by $\sim$~1~eV ($\Delta^{we}$)
\cite{jain2011quasiparticle}. These values should be the same as the CTL is a
thermodynamic quantity and should not depend on the path we chose to calculate
it.  Once we include the electrostatic correction to the E$^{QP}$ values
involving a charged defect calculation we can see the $\Delta$ values decreases
dramatically \cite{jain2011quasiparticle}.  However we find that the $\Delta$
values with including electrostatic correction are still large and the CTL has
significant cell size dependence. The remaining discrepency is due to the
choice of dielectric constants used in the electrostatic corrections.  The
electrostatic correction performed here are calculated using $\epsilon_{elec}$.
As discussed previously, by using $\epsilon_{elec}$ we assume the charges
localized at defect sites are interacting though electronically screened
Coulomb interaction only and lattice screening ($\epsilon_{latt}$) does not
play any role. But during the process of structural relaxation the surrounding
atoms move to screen the defect charges.  Therefore, in any finite supercell
size we have some lattice screening effects. Moreover we are assuming there is
no electrostatic corrections required for E$\mathrm{^{relax}}$, which we show,
in the following section, is not necessarily the case.

\begin{table}[h!]
\begin{center}
\begin{tabular}{ | c| c| c| c| c| c| c| c| c|}
\hline
CT levels & $\mathrm{E^{relax}_{q-1}}$ & $\mathrm{E^{relax}_{q}}$ & $\mathrm{E^{QP}_{P1}}$ & $\epsilon^{q/q-1}$ & $\mathrm{E^{QP}_{P2}}$ & $\epsilon^{q/q-1}$ & $\Delta$ & Mean\\
\hline
\hline
$\epsilon^{+1/0}$ & -0.12 & -1.09 & 2.87 & 3.84 & 2.94 & 3.9 & 0.07 & 3.87 \\
\hline
$\epsilon^{+2/+1}$ & -1.09 & -3.79 & -0.14 & 2.56 & -0.09 & 2.61 & 0.05 & 2.58\\
\hline
\end{tabular}
\caption{ \label{tab4}  Charge transition levels calculated with
3$\times$3$\times$1 supercell using Eq.~\ref{eqn5} with including electrostatic
corrections for both E$\mathrm{^{QP}}$ and E$\mathrm{^{relax}}$}
\end{center}
\end{table}

To address the issues mentioned earlier and to eliminate the remaining error in
CTLs following P1 and P2 we calculate the CTLs using Eq.~\ref{eqn5}. In these
calculation the quasiparticle transition energies are calculate at the
unrelaxed structures. As a result one can use $\epsilon_{elec}$ to perform
electrostatic corrections for E$\mathrm{^{QP}}$. While in principle one can
construct an electrostatic model for the correction to E$\mathrm{^{relax}}$, we
have computed it by extrapolating the relaxation energies from 4 supercell
sizes. The E$^{\mathrm{relax}}$ value extrapolated to infinite system size is
used to correct the E$\mathrm{^{relax}}$ for the finite cell sizes. In
Table.~\ref{tab4} we show charge transition levels calculated with
3$\times$3$\times$1 supercell using Eq.~\ref{eqn5} using IP (P1) as well as EA
(P2) values of unrelaxed system with defect at respective charge states. We
have performed the electrostatic corrections for E$\mathrm{^{QP}}$ using FNV
scheme. We find that the $\Delta$ values are now less than 0.1 eV. Moreover in
Table.~\ref{tab5} we show that the CTL values calculated in this way has a
much smaller cell size dependence ($<$ 0.15 eV). Calculating CTL using
Eq.~\ref{eqn5} provides a very efficient way to perform electrostatic
corrections and to obtain more accurate results than using Eq.~\ref{eqn3} or
Eq.~\ref{eqn4}. 

With modified DFT+GW formalism our calculation shows that the oxygen vacancy
has two donor levels within the gap of $\alpha$-alumina. It has a transition
level (+2/+1) at 2.5 eV and a (+1/0) level at 3.8 eV above the VBM.

\begin{table}[h!]
\begin{center}
\begin{tabular}{| c| c| c| c| c|}
\hline
Cell size  & 2$\times$2$\times$1  & 3$\times$3$\times$1 & $\Delta$ & Mean \\
\hline
\hline
$\epsilon^{+1/0}$ & 3.77 & 3.9 & 0.13 & 3.83 \\
\hline
$\epsilon^{+2/+1}$ & 2.42 & 2.56 & 0.14 & 2.49 \\
\hline
\end{tabular}
\caption {\label{tab5} Charge transition levels calculated with
2$\times$2$\times$1 and 3$\times$3$\times$1 supercell using Eq.~\ref{eqn5} with
including electrostatic corrections for both E$\mathrm{^{QP}}$ and E$\mathrm{^{relax}}$}
\end{center}
\end{table}

The optical absorption of F-center in $\alpha$-alumina involves excitation from
$^1S$-like ground state to a singlet excited state, $^1P$
\cite{evans1991optical,lee1979luminescence}. The system then relaxes to a
triplet excited state ($^3P$) using a non-radiative relaxation process.  The
emission occurs when the system goes through a $^3P$ $\rightarrow$ $^1S$
transition. As this is a spin-forbidden transition the lifetime of this process
is very high ($\sim$ 36 msec) \cite{lee1979luminescence} and results in
photoluminesence. Experimental studies of the F-center in alumina have found
the absorption and emission peaks at 6.1 eV and 3 eV respectively. 

To study F-center optical absorption we start from the neutral defect
supercell. As discussed earlier the ground state in this case has a
doubly occupied defect state at 2.55 eV from the VBM (Fig.~\ref{fig3}).
To study the singlet transition ($^1S$ $\rightarrow$ $^1P$)  we perform a BSE
calculation including both the direct ($K^d$) and exchange ($K^x$)
contributions in the electron-hole interaction kernel. We find the lowest
energy exciton at 6.2 eV.  This agrees very well with the experimental
absorption peak at 6.1 eV \cite{lee1979luminescence}.  Without electron-hole
interaction this absorption peak would have been at 6.55 eV, indicating a
strong excitonic effect with a exciton binding energy of $\sim$ 0.35 eV. This
exciton binding energy is very close to the bulk $\alpha$-alumina value (0.37 eV). 

\begin{figure}[h!]
\setlength{\unitlength}{0.1\textwidth}
\begin{picture}(5,4)
\put(0,0){\includegraphics[height=7cm] {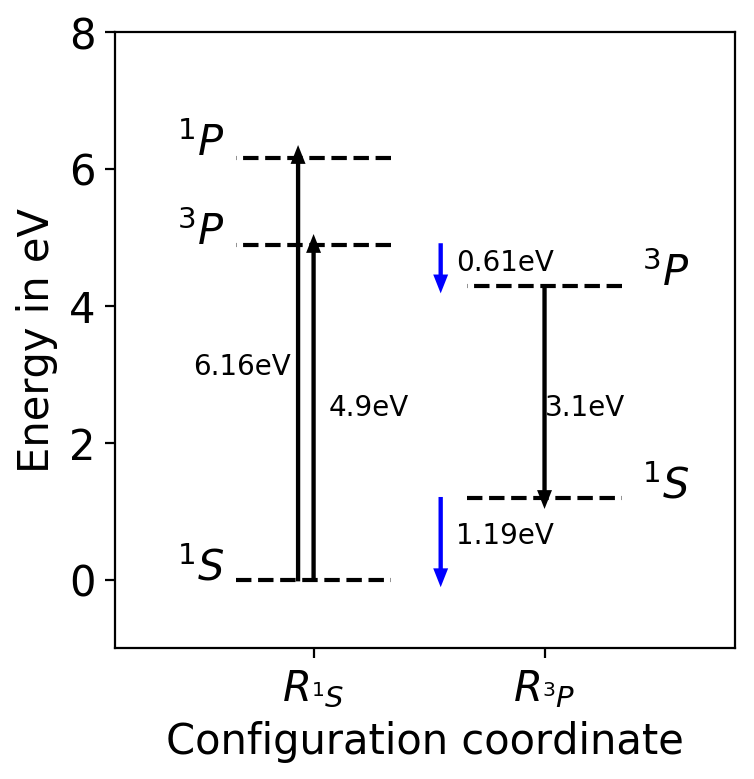}}
\end{picture}
\caption{Energy-level scheme for absorption and emission processes associated
with F-center in $\alpha$-alumina}
\label{fig4}
\end{figure}

If the absorption ($^1S$ $\rightarrow$ $^3P$) and emission ($^3P$ $\rightarrow$
$^1S$) energies are denoted by E$_{ab}$ and E$_{em}$ respectively.  It is
evident that they are related as, 

\begin{equation}
\begin{split}
E_{ab} & = E_{em}+(E^{^3P}[R_{^1S}]-E^{^3P}[R_{^3P}])+(E^{^1S}[R_{^3P}]-E^{^1S}[R_{^1S}]) \\
       & = E_{em}+E^{^3P}_{relax}+E^{^1S}_{relax}   
\end{split}
\label{eqabs}
\end{equation}

where $E^{^3P}_{relax}$ is the triplet excited state relaxation energy and
$E^{^1S}_{relax}$  is the ground state relaxation energy.  The
$E^{^3P}_{relax}$ is the total energy difference, when the system is in triplet
excited state, between all the atoms at the $^1S$ equilibrium structure
($\mathrm{\left\{{\bf R}_{^1S}\right\}}$)  and it's own equilibrium structure.
Similarly, $E^{^1S}_{relax}$ is the the total energy difference, when the
system is in singlet gound state, between all the atoms at the $^1S$
equilibrium structure ($\mathrm{\left\{{\bf R}_{^3P}\right\}}$) and it's own
equilibrium structure.

The emission process involves  $^3P$ $\rightarrow$ $^1S$ transition.  To study
this process we calculated the ``dark'' triplet solution of the BSE with the
system at the equilibrium structure of the ground state. Within BSE, triplet
solutions are found by making the exchange contribution ($K^x$) in the
electron-hole kernel to be zero. We found that the lowest energy required for this
transition is at 4.9 eV. This energy can be interpreted as $E_{ab}$ in
Eqn.~\ref{eqabs}. We then calculate excited state relaxation energy
($E^{^3P}_{relax}$) using constrained-DFT as dicussed earlier and  we find it
to be 0.61 eV. The ground state relaxation energy ($E^{^1S}_{relax}$) is then
computed by finding the excited state equllibrium structure ($R_{^3P}$) from
constrained-DFT calculation. We find this value to be 1.18 eV. Using these
results in Eqn.~\ref{eqabs} we find that the emission process should be at 3.1 eV.
This is in excellent agreement with the experimental photoluminescence emission
peak \cite{lee1979luminescence}.  In Fig~\ref{fig4} we explain the absorption
and emission process and also show all the excitation and relaxation energies
calculated for absorption and emission of F-center in $\alpha$-alumina.   
 
\section{Conclusion}

We have used first-principles methods to study electronic and optical
properties of $\alpha$-alumina. Using DFT+GW formalism we calculate the
thermodynamic charge transition levels for oxygen vacancies in this material.
We propose a modified version of this formalism which can be used to perform
electrostatic correction more efficiently and therefore provide more accurate
CTL values. We find that oxygen vacancy in this material has deep donor
levels, (+2/+1) at 2.5 eV and a (+1/0) level at 3.8 eV above the VBM.  We also
study the optical absorption and emission process of neutral oxygen vacancy in
$\alpha$-alumina (F-center) using GW-BSE methodology. For the photolumiscence
emission process, we use constrained-DFT to simulate ground and excited state
relaxation processes. Our calculation suggests that F-center in this material
will have absorption and emission peak near 6.2 eV and 3.1 eV respectively,
which is in very good agreement with experimental results.

\bibliography{alumina} 
\end{document}